\documentclass[twocolumn,prb,aps,showpacs,superscriptaddress]{revtex4-1}
\usepackage{graphicx}
\usepackage[dvipdfm]{hyperref} \newcommand{\hm}{h_{mw}}

\usepackage[usenames,dvipsnames]{pstricks} 
\usepackage{pst-plot} 
\usepackage{epsfig} 
\usepackage{pst-grad} 
\usepackage{pst-plot} 
\usepackage{pstricks-add}

\usepackage{lmodern} 
\newcommand{\mycol}{1} \newcommand{\beq}{\begin{equation}}
\newcommand{\eeq}{\end{equation}}

\usepackage{amsmath} \usepackage{ulem} 

\begin{document} 

\title{Resonant single and multi-photon coherent transitions in a detuned
regime}

\author{S. Bertaina} \email{sylvain.bertaina@im2np.fr} \affiliation{IM2NP-CNRS
(UMR 7334) and Aix-Marseille Universit\'{e}, Facult\'{e} des Sciences et
Techniques, Avenue Escadrille Normandie Niemen - Case 162, F-13397 Marseille
Cedex, France.}

\author{M. Martens} \affiliation{Department of Physics and The National High
Magnetic Field Laboratory, Florida State University, Tallahassee, Florida
32310, USA}

\author{M. Egels} \affiliation{IM2NP-CNRS (UMR 7334) and Aix-Marseille
Universit\'{e}, Facult\'{e} des Sciences et Techniques, Avenue Escadrille
Normandie Niemen - Case 162, F-13397 Marseille Cedex, France.}

\author{D. Barakel} \affiliation{IM2NP-CNRS (UMR 7334) and Aix-Marseille
Universit\'{e}, Facult\'{e} des Sciences et Techniques, Avenue Escadrille
Normandie Niemen - Case 162, F-13397 Marseille Cedex, France.}

\author{I. Chiorescu} \affiliation{Department of Physics and The National High
Magnetic Field Laboratory, Florida State University, Tallahassee, Florida
32310, USA}

\date{\today}%

\begin{abstract}
We performed quantum manipulations of the multi-level spin system S=5/2 of a
Mn$^{2+}$ ion, by means of a two-tone pulse drive. The detuning between the
excitation and readout radio frequency pulses allows one to select the number of
photons involved in a Rabi oscillation as well as increase the frequency of this
nutation. Thus detuning can lead to a resonant multi-photon process. Our
analytical model for a two-photon process as well as a numerical generalization
fit well the experimental findings, with implications in the use of multi-level
spin systems as tunable solid state qubits.
\end{abstract}


\maketitle

Quantum properties of electronic spins can be controlled due to their relatively
long coherence times. This is mostly achieved by spin dilution in a non-magnetic
matrix
\cite{Nellutla2007,Dutt2007,Bertaina2007,Ardavan2007,Bertaina_NatLett_2008,Bertaina2009},
leading to coherent Rabi oscillations up to room
temperature\cite{Jelezko2004,Nellutla2007}. Fundamental and technological
advances can be achieved by strongly coupling spins with photons in a
cavity\cite{Chiorescu2010,Kubo2010,Schuster2010}, leading to a new type of
hybrid quantum memory\cite{Blencowe2010}. Recently, we proposed a tunable
multi-level system as a potential candidate for multi-qubit
implementation\cite{Bertaina2009,Bertaina2011} which could be used to implement
Grover's algorithm\cite{Leuenberger2001,Leuenberger2003,Grace2006} for instance.
We describe in this work a two-tone experiment using the $S=5/2$ spin of
Mn$^{2+}$ ions diluted in MgO. This technique allows us to study multi-photon
dynamics in and out of resonance by detuning the two radio-frequencies of the
excitation and readout pulses. Interestingly, the detuning can actually be used
to bring into resonance levels separated by exactly two or more photons. The
implication is that the multi-level electron spin dynamics can now be controlled
using any number of photons, anywhere within the dressed state energy diagram.
In addition, detuning Rabi frequencies increases the nutation frequency, which
gives access to fast Rabi nutation speeds\cite{Yoneda2014,Chesi2014}.

\begin{figure} \centering
	\includegraphics[bb=55 505 312 741,clip, width=0.8\mycol\columnwidth]{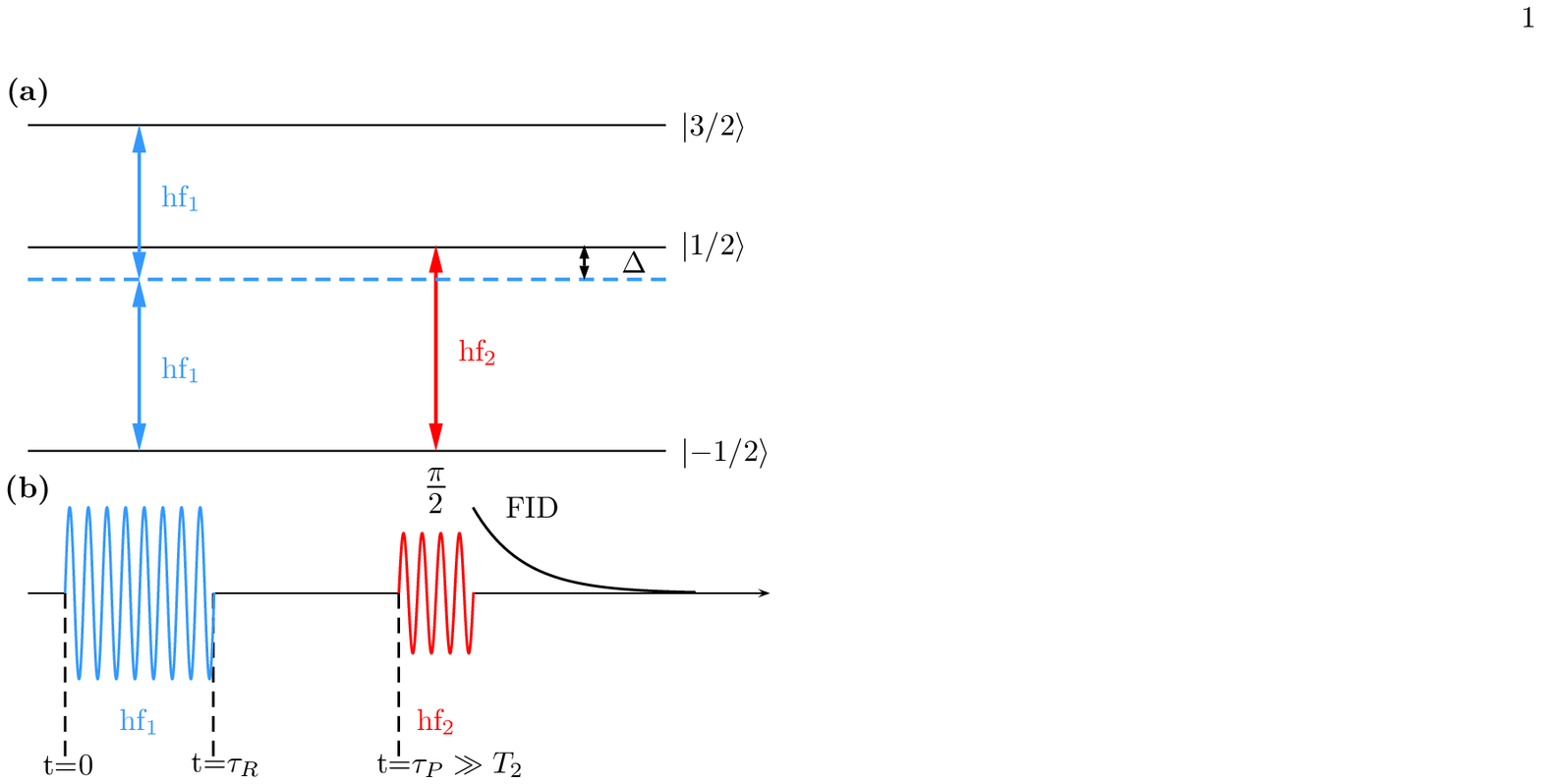} 
	\caption{(Color online). Representation of a two-photon pump and probe process.
		(a) Energy levels of Mn$^{2+}$ (see Eq.~\ref{eq:3}) for the subgroup
		$\left|\pm1/2\right>$ and $\left| 3/2\right>$. The arrows indicate the photon
		absorptions. The dashed line is a virtual level equidistant between
		$\left|-1/2\right>$ and $\left|3/2\right>$. (b) Pulse sequence: at $t=0$ the
		microwave of frequency $f_1$ irradiates the spins and induces a Rabi nutation
		lasting up to t$=\tau_R$. To probe the population at the end of the first pulse,
		a second pulse at frequency $f_2$ in resonance with the transition between
		$\left| -1/2\right >$ and $\left| 1/2\right >$ induces a $\pi/2$ rotation and
		thus an FID signal in spectrometer.}
	\label{fig:1} 
\end{figure}

The $S=5/2$ system of Mn$^{2+}$ ions diluted in MgO has been extensively
studied by EPR since it was first considered as model for crystal field theory
\cite{Low1957}. The Mn$^{2+}$ ions are substituted for Mg$^{2+}$ and have a
cubic symmetry $F_{m\bar{3}m}$ (lattice constant 4.216~{\AA}) ensuring that the
spins see an almost isotropic crystalline environment. All the parameters of the
crystal field and hyperfine interactions are known by independent studies.  The
spin Hamiltonian resolved by an electron paramagnetic resonance experiment (EPR)
is given by \cite{Low1957,Smith1968}:

\begin{eqnarray}
\label{eq:1}
H&=&a/6\left[
{S_x^4+S_y^4+S_z^4-S(S+1)(3S^2-1)/5} \right]\\
&&+\gamma\vec{H}_0\cdot\vec{S}-A\vec{S}\cdot\vec{I}+\gamma\vec{h}_{mw}\cdot\vec{S}
\cos(2\pi ft) \nonumber 
\end{eqnarray}
where $\gamma=g\mu_B/h$ is the gyromagnetic ratio ($g=2.0014$ the $g$-factor,
$\mu_B$ Bohr's magneton and $h$ Planck's constant), $S_{x,y,z}$ are the spin
projection operators, $\vec{S}$ is the total spin, $a=55.7$~MHz is the
anisotropy constant, $A = 244$~MHz is the hyperfine constant of $^{55}$Mn
($I=5/2$), $h_{mw}$ and $f$ represent the MW amplitude and frequency
respectively, and $\vec{H}_0$ is the static field ($\vec{H}_0\perp
\vec{h}_{mw}$). The applied static field ensures a Zeeman splitting of $\gamma
H_0\approx f\sim9$~GHz, much stronger than all other interactions of
Eq.(\ref{eq:1}). This implies that (i) $\vec{H}_0$'s direction can be
approximated as the
quantization axis and (ii) coherent microwave driving is confined between levels
of same nuclear spin projection $m_I$ (see also refs.
\onlinecite{Bertaina2009,Hicke2007,Schweiger2001}). Consequently, the hyperfine
interaction generates a constant field shift for all levels and it will be
dropped from further analytical considerations (although it is part of the full
numerical simulations).

Let us consider a quantum system with six states $| S_z\rangle$,
$S_z$=$\left\{-5/2, -3/2, -1/2, 1/2, 3/2, 5/2 \right\}$, irradiated by an
electromagnetic field. The spin Hamiltonian of the system is:
\begin{equation}
\label{eq:2}
\mathcal{H}=\hat{E}+\hat{V}(t)=\sum_{S_z=-5/2}^{5/2} E_{S_z}|S_z\rangle\langle
S_z|+\hat{V}(t), 
\end{equation}
with $E_{Sz}$ the static energy levels,
$\hat{V}(t)=\frac{\gamma}{2}h_{mw}(\hat{S}_++\hat{S}_-)\cos\left(2\pi
ft\right)$, $S_+/S_-$ the raising/lowering operators. Note that, contrary to our
previous studies \cite{Bertaina2009a,Bertaina2011} where $f$ was fixed to be in
resonance with 1/2 and -1/2 levels, here $f$ is a free parameter. In a cubic
symmetry and in he first order perturbation theory ($a \ll H_0$), the
static energy levels are given by:
\begin{eqnarray} 
\label{eq:3}
\nonumber E_{\pm 5/2}&=&(\pm 5/2) \gamma H_0+(1/2) p
a+\mathcal{O}(a^2)\\ E_{\pm 3/2}&=&(\pm 3/2) \gamma H_0-(3/2) p
a+\mathcal{O}(a^2)\\ E_{\pm 1/2}&=&(\pm 1/2) \gamma H_0+ p a+\mathcal{O}(a^2)
\nonumber
\end{eqnarray}
where\cite{Low1957} $p=1-5\sin^2\theta+\frac{15}{4}\sin^4\theta$ with $\theta$
the angle between $\vec{H}_0$ and the $c$ axis $[001]$. Since $H_0\gg \hm$, we
can use the rotating wave approximation (RWA) to make Eq.~(\ref{eq:2}) time
independent. We apply the unitary transformation $U(t)=\exp(-i2\pi f \hat{S}_z
t)$ and the Hamiltonian (\ref{eq:2}) becomes \cite{Hicke2007, Leuenberger2003}:%
\begin{equation} \label{eq:5} 
\mathcal{H}_{RWA}=U\mathcal{H}
U^\dag+i\hbar\frac{\partial U}{\partial t}U^\dag= 
\end{equation}

\begin{widetext} 
\begin{equation} 
=\left(\begin{matrix}
\frac{1}{2}pa-\frac{5}{2}\Delta  & \frac{\sqrt{5}}{2} V & 0 & 0 & 0 & 0\\
\frac{\sqrt{5}}{2} V&-\frac{3}{2} p a -\frac{3}{2}\Delta & \sqrt{2} V  & 0 & 0
& 0\\ 0 &\sqrt{2} V & pa-\frac{1}{2}\Delta & \frac{3}{2}V & 0 & 0\\ 0 &  0  &
\frac{3}{2}V  & pa+\frac{1}{2}\Delta  & \sqrt{2} V  & 0\\ 0 &  0  & 0 & 
\sqrt{2} V  & -\frac{3}{2} p a +\frac{3}{2}\Delta & \frac{\sqrt{5}}{2} V\\ 0 &0
& 0 & 0 &  \frac{\sqrt{5}}{2} V  & \frac{1}{2}pa+\frac{5}{2}\Delta \\
\end{matrix}\right) \label{eq:6} 
\end{equation}
\end{widetext} 
where $V=\gamma\hm/2$ and $\Delta$ is the detuning parameter defined by
$\Delta=f- (E_{1/2}-E_{-1/2})$. By diagonalization, the eigenenergies $E_n/pa$
of the dressed states are calculated as a function of $V$ and $\Delta$. The Rabi
frequency is the energy difference between two consecutive dressed states.
Depending on the value of $\Delta$, one can probe the ``in resonance one photon
process'' ($\Delta=0$), the ``detuning regime one photon process'' ($\Delta \neq
0$) and the ``multiphoton process''. The case $\Delta=0$ has been reported in a
previous study \cite{Bertaina2011}. An example for the latter case, for which we
discuss experimental evidence below, is $\Delta= \pm5pa/4$ when two two-photon
resonances occur.

In a typical pulsed EPR experiment there is only one frequency at a time
available and the pulse sequence is composed of two parts: the spin manipulation
and the probe sequence. Since the later has to be at a frequency $f$ in
resonance with a one photon transition, the experiment is restricted to be at
$\Delta = 0$. In the current work, we use two microwave sources which allows us
to manipulate any dressed state by using the first source $f_1$ and then probe
the induced variation of populations using the second source $f_2$. A
representation of the pulse sequence used for coherent manipulation in the case
of a two-photon process is described in Fig.~\ref{fig:1}. To simplify the
explanation only 3 out of 6 levels are shown. At $t=0$, an oscillating field of
frequency $f_1$ irradiates the system for a time $\tau_R$. When the detuning
$\Delta$ is such that the levels $\left| -1/2\right>$ and $\left| 3/2\right>$
are separated by exactly $2hf_1$, a two-photon coherent transition is induced.
To probe the populations of levels, a second pulse at frequency $f_2$ resonant
with the transition between $\left| -1/2\right>$ and $\left| 1/2\right>$ is sent
at a time $t=\tau_p>>T_2$ but smaller than $T_1$, where $T_{1,2}$ are the spin
relaxation and decoherence times respectively. This second pulse induces a free
induced decay (FID) with an intensity proportional to the population difference
$\sigma_{-1/2}-\sigma_{1/2}$ via a $\pi/2$ rotation in the sub-space $S_z=\pm
1/2$.

The Rabi oscillation measurements were performed using a conventional Bruker
Elexsys 680 pulse EPR spectrometer working at about 9.6~GHz. The second
frequency source is provided by the ELDOR bridge of the spectrometer. To verify
the reproducibility of the experiments, two different spectrometers using two
different resonators, a dielectric (MD-5) and a split coil (MS-5) were used. The
power-to-field conversion rate of the resonators is dependent of the frequency
$f_1$ and it has been calibrated for all the frequencies used in this study
using DPPH, an isotropic $S=1/2$ system. The sample is a
(3$\times$3$\times$1)-mm$^3$ single crystal of MgO doped by a small amount of
Mn$^{2+}$.

A Rabi measurement consists of recording the FID intensity, as explained above,
as a function of pulse length $\tau_R$. Such Rabi oscillations are acquired for
different excitation frequencies $f_1$, around the main one-photon Zeeman
resonance $f_2$. Other important experimental parameters are the sample
temperature and external field orientation (parameter $p$). In the current work,
the temperature was set at 50~K which provides a long enough Rabi coherence time
($\sim 1\mu$s), while keeping the relaxation time sufficiently short to ensure a
fast acquisition time \cite{Bertaina2009,Bertaina2011}. Thus, the waiting time
between the pump and probe pulses was set at 3~$\mu$s, larger than the
decoherence time, but shorter than the relaxation time. The external field was
oriented along the $[111]$ axis of the crystal, ensuring a sizable anharmonicity
of the six level system ($p=-2/3$) allowing the creation of virtual levels
depicted by the dashed line in Fig.~\ref{fig:1}b. It is important to note that
the results described below are achievable in principle, for any orientation of
the external field, the level anharmonicity being a tuning parameter $V/(pa)$ in
such a multi-level quantum system.   
\begin{figure}
	\includegraphics[width=\mycol\columnwidth]{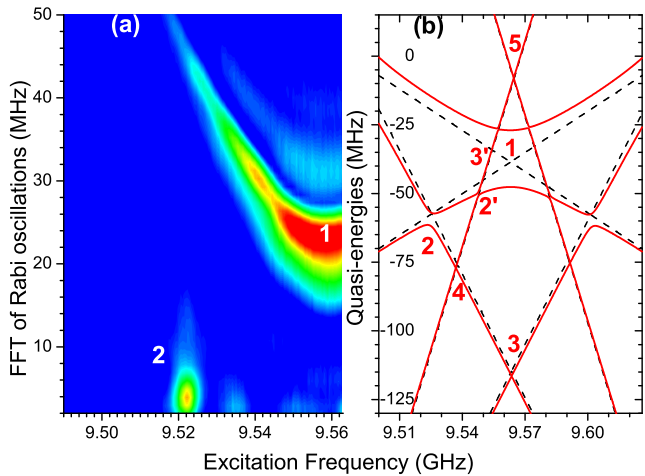}
	\caption{(Color online). (a) Fourier transform of detuned Rabi oscillations
		measured at medium microwave power, $h_{mw}=0.5(3)$~mT. The one-photon branch,
		marked with ''1" can reach tens of MHz. An off-resonance two-photon branch,
		marked with ''2", is also visible. The color map is in arbitrary units. (b)
		Quasi-energies of Hamiltonian~\ref{eq:6} calculated for low (dashed lines) and
		medium (continous lines) microwave powers. The one-photon Rabi splitting (1) is
		excited when $f_1=f_2$ while the two-photon transition (2) is at a detuned
		location given by Eq.~\ref{eq:f2}. }
	\label{fig:FIG_FFTdressed} 
\end{figure} 
\begin{figure}
	\includegraphics[width=\mycol\columnwidth]{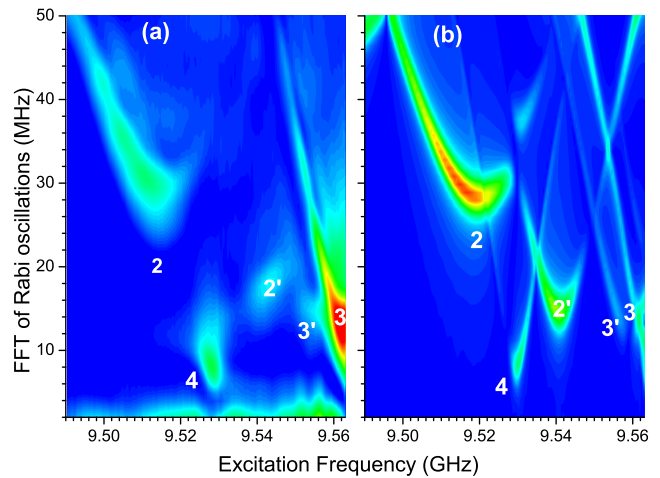} 
	\caption{(Color online). Measured (a) and simulated (b) Fourier transform of
		Rabi oscillations in the case of high microwave power ($h_{mw}=1.7(8)$~mT). The
		very good agreement allows identification of each Rabi splitting (shown in
		Fig.~\ref{fig:FIG_FFTdressed}): two-photon (2 and 2'), three-photon (3 and 3')
		and four-photon coherent rotations (4). The color map is in arbitrary units.}
	\label{fig:FIG_highpower}
\end{figure}
\begin{figure}
	\includegraphics[width=\mycol\columnwidth]{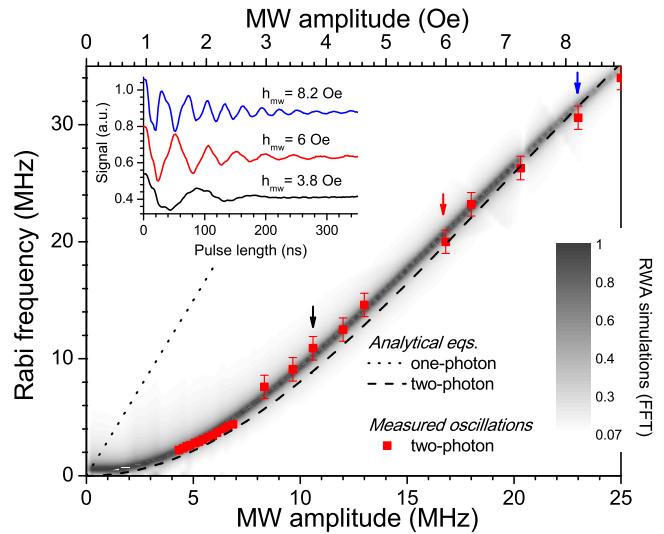} 
	\caption{(Color online).  Microwave field dependence of two-photon coherent
		Rabi oscillations. The contour plot of the background is obtained via numerical
		simulations using the RWA (shade intensity in arbitrary units). The dashed line
		is given by Eq.~(\ref{eq:2pRabiFreq}) and it is in excellent agreement with the
		experimental values of the Rabi frequency (red squares). For comparison, the
		linear dependence of the one-photon Rabi frequency is shown by a dotted line.
		Two-photon Rabi coherent oscillations are shown (insert) for three values of the
		microwave field, as indicated by small vertical arrows in the main panel.} 
	\label{fig:FIG_2photons}
\end{figure}
\begin{figure}
	\includegraphics[width=\mycol\columnwidth]{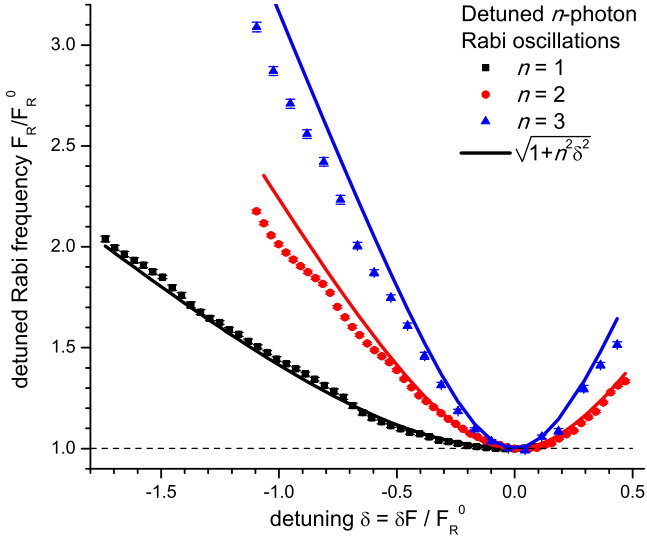} 
	\caption{ (Color online). Acceleration of $n$-photon Rabi frequencies as a
		function of detuning. Experimental data is showing by black squares
		(one-photon), red dots (two-photon) and blue triangles (three-photon Rabi
		splittings), while the continuous lines are calculated from Eq.~\ref{eq:10}.  } 
	\label{fig:FIG_nphotons}
\end{figure}

After a Rabi oscillation is recorded, a Fourier transform indicates the Rabi
frequency and its decay properties. An important aspect of our study relies on
the possibility of detuning the excitation frequency $f_1$, as shown in the
contour plot of Fig.~\ref{fig:FIG_FFTdressed}a. For moderate powers
[$h_{mw}=0.5(3)$~mT], one observes a fast one-photon branch (marked with ``1'')
reaching Rabi flops of several tens of MHz. The branch shows a typical
square-root law as a function of detuning~\cite{Cohen-Tannoudji1978} (see also
Fig.~\ref{fig:FIG_nphotons}). This technique allows for a significant speed-up
of electronic or nuclear spin Rabi
frequencies\cite{Rohr2014,Avinadav2014,Yoneda2014} which are usually much slower
than those achievable in superconducting qubits\cite{Chiorescu2004}. The
quasi-energies calculated as eigen-energies of Hamiltonian~(\ref{eq:6}) are
shown in Fig.~\ref{fig:FIG_FFTdressed} as a function of $f_1$. Crossings of the
dashed lines (low power case) indicate the location of resonances by
cross-overs, e.g. the one-, three- and five-photon resonances at $\Delta=0$
(discussed in our previous studies  \onlinecite{Bertaina2009,Bertaina2011}). The
continuous lines represent the eigen-energies calculated for the same power as
the experimental data in Fig.~\ref{fig:FIG_FFTdressed}. One observes the large
one-photon splitting (1) which can be further accelerated by detuning (here,
$f_1<f_2$). A two-photon splitting of $\sim4$~MHz is visible as well (actual
two-photon Rabi oscillations are shown in Fig.~\ref{fig:FIG_2photons}).

The Rabi splittings between consecutive quasi-energies are strongly dependent on
the microwave power. A Fourier transform contour plot for the high power regime
[$h_{mw}=1.7(8)$~mT] is shown in Fig.~\ref{fig:FIG_highpower}: the panel (a)
shows experimental data while panel (b) shows numerical simulations based on
exact diagonalization of Hamiltonian~\ref{eq:1} after the RWA unitary
transformation~\ref{eq:5}. One observes two two-photon coherent rotations (2 and
2'), two three-photon processes (3 and 3') as well as a four-photon detuned
oscillations. The positions of these transitions as a function of detuning is
visible in Fig.~\ref{fig:FIG_FFTdressed}b. The very good agreement between
experiment and simulation allows the identification of all Rabi frequencies, and
thus a predictable method of tuning level superposition between various $S_z$
states of the Mn spin. Moreover, the possibility of frequency detuning between
the control pulse $f_1$ and the $\pi/2$ read-out pulse $f_2$ allows for
selection of certain Rabi splittings without changing the external magnetic
field. In the following, we demonstrate this protocol in the case of detuned
two-photon coherent spin manipulation.

As discussed above, for $\Delta=-5pa/4$, one brings in resonance the levels
$E_{1/2}$ and $E_{-3/2}$. Thus, the excitation frequency is given by:
\begin{equation} \label{eq:f2}
f_1=\frac{1}{2}(E_{1/2}-E_{-3/2}) 
\end{equation} 
and Hamiltonian~(\ref{eq:6}) becomes
\begin{equation}
\mathcal{H}_{RWA} = \begin{pmatrix} \frac{29}{8}pa & \frac{\sqrt{5}}{2} V &
0&0&0&0 \\ \frac{\sqrt{5}}{2} V &\frac{3}{8}pa &\sqrt{2} V&0&0&0  \\ 0
&\sqrt{2} V &\frac{13}{8}pa& \frac{3}{2}V & 0 & 0  \\ 0 & 0 & \frac{3}{2}
V&\frac{3}{8}pa & \sqrt{2} V &0 \\ 0 & 0 & 0 & \sqrt{2}V
&-\frac{27}{8}pa&\frac{\sqrt{5}}{2} V \\ 0 & 0 & 0 & 0 & \frac{\sqrt{5}}{2}V
&-\frac{21}{8}pa \end{pmatrix} 
\end{equation}
showing the coupling between the diagonal elements $3pa/8$ via a two-photon
process.

Taking $\tilde{h} = V/pa$ and restricting the analysis to low powers (keeping
only terms quadratic in $\tilde{h}$), one finds that there is one eigenvalue
$\epsilon=3/8$. By looking for another eigenstate $\epsilon'$ close to 3/8
($F_{Rabi}^{2photon}=\epsilon-\epsilon'\ll1$), one can solve analytically for
the Rabi frequency associated with this two-photon process:
\begin{equation} \label{eq:2pRabiFreq} 
F_{Rabi}^{2photon}=\frac{1902 \tilde{h}^2}{585+713\tilde{h}^2} 
\end{equation} 
as a function of microwave power. This analytical relationship is in excellent
agreement with experimental data (red squares), as shown in
Fig.~\ref{fig:FIG_2photons}. The contour plot of the background is calculated
using full diagonalization of time-dependent Schr\"odinger equation, while the
dashed line is given by Eq.~(\ref{eq:2pRabiFreq}), with no fit parameters. The
linear microwave field dependence of a one-photon Rabi frequency is given as
dotted line, as a comparison. The insert shows actual two-photon Rabi
oscillations for three values of the microwave field, as indicated by the small
vertical arrows in the main panel. These measurements are done at the two-photon
resonance (marked with ``2'' in Fig.~\ref{fig:FIG_FFTdressed}b) by properly
detuning $f_1$ by respect to $f_2$.

One can also study the effect of the detuning away from a particular Rabi
resonance. We record spin oscillations for value of $f_1$ around Rabi splittings
corresponding to one-, two- and three-photon processes (marked with ``1'',``2''
and ``3'' respectively in Fig.~\ref{fig:FIG_FFTdressed}b). Their Fourier
transform give the values of the Rabi splittings, as plotted in
Fig.~\ref{fig:FIG_nphotons}. The obtained frequencies follow very well a
generalized Rabi formula for $n$-photons, given by:
\begin{equation}
\label{eq:10}
F_R=\sqrt{(F_R^0)^2+n^2\delta F^2}
\end{equation}  
where $F_R^0$ is the Rabi frequency at resonance, $n$ is the number of photons
responsible for the coherent drive and $\delta F$ is the detuning away from a
Rabi resonance. For instance, in the case of the two-photon coherent
oscillations, $n=2$, $F_R^0=F_{Rabi}^{2photon}$ of Eq.~(\ref{eq:2pRabiFreq}) and
the detuning is the shift of $f_1$ away from the splitting marked ''2" in
Fig.~\ref{fig:FIG_2photons}. For one-photon, the above equation is reduced to
the well-known Rabi formula\cite{Cohen-Tannoudji1978}. This observed
acceleration of the Rabi flops due to detuning can be an important tool to
achieve fast operation of spin qubits\cite{Yoneda2014}. 

In conclusion, we demonstrate a pulse technique allowing for coherent operation
of multi-photon Rabi oscillations through detuning. Two microwave pulses are
detuned such that one excites with a specified number of photons while the other
one provides the readout at the single photon resonance. The technique allows
for a significant speed-up of Rabi nutation by detuning the one or multi-photon
coherent rotation. The provided analytical and numerical methods allows for a
well defined control and tuning of the spin dynamics in this multi-level solid
state system.

This work was supported by NSF Grant No. DMR-1206267, CNRS-PICS CoDyLow and
CNRS's research federation RENARD (FR3443) for EPR facilities. The NHMFL is supported
by Cooperative Agreement Grant No. DMR-1157490 and the state of Florida.

\bibliographystyle{apsrev4-1}
\bibliography{Biblio}

\end{document}